# Transference& Retrieval of Pulse-code modulation Audio over Short Messaging Service

Muhammad Fahad Khan
Federal Urdu University of
Arts, Science and
Technology.
G7/1, Islamabad, Pakistan.

Saira Beg
COMSATS Institute of
Information and Technology,
Park Road, Chak Shahzad,
Islamabad, Pakistan.

## ABSTRACT
The paper presents the method of transferring PCM (Pulse-Code Modulation) based audio messages through SMS (Short Message Service) over GSM (Global System for Mobile Communications) network. As SMS is text based service, and could not send voice. Our method enables voice transferring through SMS, by converting PCM audio into characters. Than Huffman coding compression technique is applied in order to reduce numbers of characters which will latterly set as payload text of SMS. Testing the said method we develop an application using J2me platform.

## 1. INTRODUCTION
SMS service of GSM is widely used in all over the world. It has national and international roaming and also supported by other major technologies as well. SMS size is limited to 160 characters and it can only send alphanumeric text only [1]. The application extension of SMS is EMS (Extended Messaging Service), which has larger contents (text, animations, predefined sounds and images) but less supported than SMS and its all components are present in message header which will ignored in unsupported mobile phones [2].

Presented method has three major steps; first step is about converting user input into characters and second step performs compression method on those characters and third step converts the compressed characters into strings and set them as payload text of SMS. For compression, Huffman Coding method is used. Main reasons for selecting Huffman Coding are; [3, 10] it is easier than other compression method as compared to arithmetic coding, needs less execution time, performs well in case of multimedia contents etc.

In this paper we present an alternative way of sending PCM based audio messages through SMS. As SMS is text based service, so a method is developed which converts audio messages into characters. After converting lossless compression technique; Huffman coding is applied. Such compression method focuses on the frequencies of characters and then represents those frequencies and characters in tree. Lastly those characters will set as a payload text of SMS. Paper is divided into following sections. Section 2 discusses the related work, section 3 and 4 is about proposed methodology and results and lastly we conclude the paper.

## 2. RELATED WORK
In this paper, a method and equipment is provided for conveying cellular signaling and other SS7 based signaling for convergence of cellular signaling, voice and data over a packet switching network. It also enables existing cellular equipment particularly existing MSCs to convey signaling over packet switched links while minimizing changes required in cellular equipment for this purpose [4]. [5] discussed a method which allows wireless user to obtain information from World Wide Web, internet or other information source via SMS or micro browser in phone. Method uses a dialed telephone number, feature code, other dialed digits or SMS origination message to cause SMS and micro browser messages to be sent to a wireless telephone or other device. GSM-SMS is also used for data acquisition in a field. A study of feasibility on application of GSM-SMS technology over field data acquisition is provided. As GSM-SMS technology provides data transmission, data communication and control of transmission and communication. Author's presents the GSM-SMS architecture that is based on characteristics of transmission and capacity of short message, moreover a package format of short message, is also developed [6].

[7] presents the method of transferring voice using SMS over GSM network. Such method takes output of encoder card known as utterance as input, than converts those utterances as strings and put those strings as payload text of SMS. [8] is about transferring voice through SMS. In this method author converts the audio messages in strings and set those strings as a payload text of SMS. One major issue with this method is that; it generates large number of concatenated SMS.

## 3. PROPOSED METHODOLOGY
For transferring and retrieving of PCM based audio messages through SMS, install our application on a mobile phone. It takes voice messages as input from user and converts that input into a SMS; hides variant steps from the user. The variant steps are;

1. First, it stored user input in a ByteArrayOutputStream.
2. Second, it converted the signed ByteArrayOutputStream into unsigned integer array.

3. Third, unsigned integer array was converted into their respective Extended ASCII characters. But before that conversion, 256 was added in all unsigned integer array values which was ranged between 0-31 in ordered to move them up to the range of 256- 287. The main reason behind this was that, values of 0-31 of ASCII characters cannot send through SMS. Because such characters were universally reserved for specific functions; '0' represents 'null' in ASCII etc.
4. Fourth, apply lossless compression algorithm Huffman coding on Extended ASCII characters. Huffman coding algorithm focuses on the frequency of the characters; and generally frequency represents in a tree format.
5. Fifth, now convert ASCII characters into the strings and set those strings as a payload text of SMS.

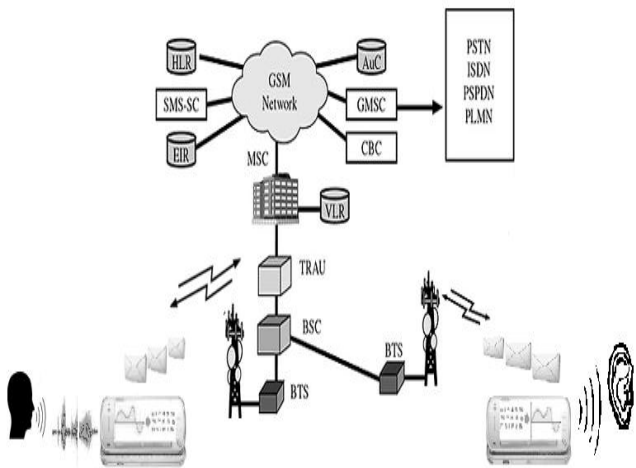

Figure 1: Existing Architecture and Proposed System interaction

Characters generated by the application could not be adjusted in one SMS; it may consume multiple numbers of SMS. Here, we used the extended SMS called Concatenated SMS [9]. Linking each SMS for one message we used indexing. For this purpose first three characters of SMS is reserved which give 0-999 indexing values. When SMS received at receiver side, our application first stored all SMS in order by using their index number than extract payload text of SMS and applies all steps in reverse in order to generate actual voice message. Figure 1 describes the proposed system interaction within the existing GSM architecture. Our system dose not requires any hardware or intermediate source while transferring the voice. It is just an application which installed in a mobile phone.

## 4. RESULTS AND DISSCUSSION

We developed an application using J2me platform and used Nokia 3110c mobile phones for testing. For results we keep track of two main factors; number of characters and number of connected SMS. Both factors are directly proportional to each other. Table one shows the basic characteristics of voice messages; three different sentences are used which have different number of word etc.

Table 1: Basic characteristics of voice messages

| Sentence 1 | This is an Audio Clip | | |
|---|---|---|---|
| Sentence 2 | The five boxing wizards jump quickly | | |
| Sentence 3 | By Jove, my quick study of lexicography won a prize | | |
| Sentence | Sentence 1 | Sentence 2 | Sentence 3 |
| Test Numbers. | 1 to 10 | 11 to 20 | 21 to 30 |
| Length (letters) | 17 | 31 | 41 |
| Length (Words) | 5 | 6 | 10 |

For experiments we used different voice sentences with different time durations; three sentences are shown in table 1. The results of experiments are shown in figure 1 a, b. we consider two cases; with and without compression. Case one is without compression and its result is shown in blue line. Where case two describes the compression technique within the application and its result is in red line.

Figure 2 a shows the compression of two cases; with or without compression. And graph is evident that Huffman Coding perform very well with the method. Similarly figure 2b represents the compression of both cases in form of connected SMS.

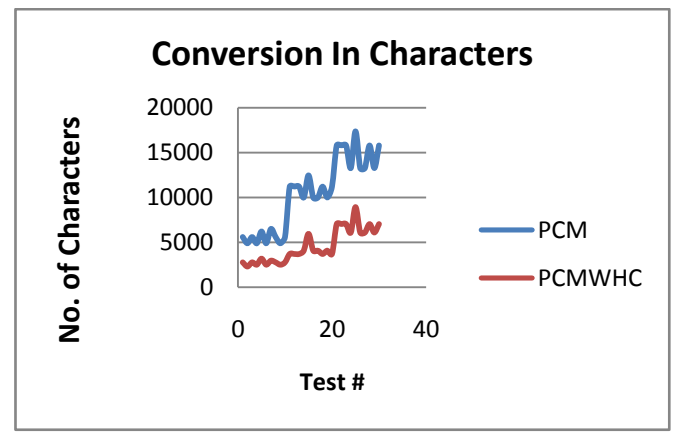

Figure 2a: Number of characters

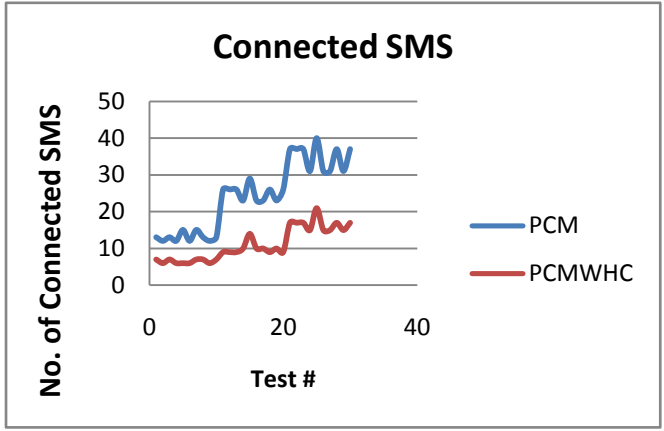

Figure 2b: Number of connected SMS

## 5. CONCLUSION

SMS is text based service and could not send voice. This paper presents the alternative way of sending voice messages using SMS over GSM network. Using this application all GSM devices can send or receive PCM based voice messages. This method is very simple and did not need any changes in existing infrastructure of GSM-SMS. In this paper we also compare the efficiency of Huffman coding method with the method and shows better results.

## 6. REFERENCES

[1] Mohd Helmy Abd Wahab, IAEng Member, Suresh P. Gopalakrishna, Ayob Haji Johari "Speed Trap Image Transfer through GSM Network " Proceedings of the World Congress on Engineering and Computer Science, San Francisco, USA, pp 231-235, 2008

[2] http://e-articles.info/e/a/title/The-Main-Protocols-used-by-Mobile-Phones-(SMS-EMS-MMS-WAP)/

[3] Eman Abdelfattah and Asif Mohiuddin, "Performance Analysis of Multimedia Compression Algorithms", nternational journal of computer science & information Technology (IJCSIT) Vol.2, No.5, pp 1-10, October 2010.

[4] Oren T. Shmulevich et al, "Convergence of Telephone signaling, voice and data over a packet switched network", United States Patent Application, Publication Number 6515985 B2, February 2003, online available: http://www.google.com.pk/patents?hl=en&lr=&vid=USPAT6515985&id=2dEOAAAAEBAJ&oi=fnd&dq=voice+over+SMS&printsec=abstract#v=onepage&q=voice%20over%20SMS&f=false

[5] Sunit Lohtia, Wilfred Martin James, Boon Chong Hwang, "System and Method for subscriber initiated information over the SMS or Micro Browser",United States Patent Application, Publication Number 6560456B1, May 2003. Online available: http://www.google.com.pk/patents?hl=en&lr=&vid=USPAT6560456&id=8QMPAAAAEBAJ&oi=fnd&dq=voice+over+SMS&printsec=abstract#v=onepage&q=voice%20over%20SMS&f=false

[6] Chwan-Lu Tseng et al, "Feasibility study on application of GSM–SMS technology to field data acquisition", Computers and Electronics in Agriculture, Volume 53, Issue 1, Pages 45-59, August 2006.

[7] Daniel L. ROTH "Voice over Short Message Service", United States Patent Application, Publication Number US2009/0017849 AI, and Origin: BOSTON, MA US, IPC8 Class: AH04Q720FI, USPC Class: 455/466, 2009. Online Available: http://www.freepatentsonline.com/20090017849.pdf

[8] M. Fahad Khan and Saira Beg, "Transferring Voice using SMS over GSM Network", Journal of Computing (JoC), pp 50-53, Volume 3, Issue 4, April 2011. ISSN-2151-9617

[9] Sun Microsystems, Inc"Wireless Messaging API (WMA) for Java™ 2 Micro Edition Reference Implementation", Version 1.0, JSR 120 Expert Groups, 2002.

Asadollah Shahbahrami, Ramin Bahrampour, Mobin Sabbaghi Rostami, Mostafa Ayoubi Mobarhan, "Evaluation of Huffman and Arithmetic Algorithms for Multimedia Compression Standards", 2011. Online Available: http://arxiv.org/ftp/arxiv/papers/1109/1109.0216.pdf